\documentclass[reprint,
superscriptaddress,
showpacs,
 amsmath,amssymb,
 aps,
]{revtex4-1}

\usepackage{mathrsfs}
\usepackage{latexsym}
\usepackage{graphicx}
\usepackage{dcolumn}
\usepackage{bm}

\begin{document}

\preprint{APS/123-QED}

\title{Quantum spin transport and dynamics through a novel F/N junction}

\author{Hua Li}
\email[E-mail: ]{huali@bc.edu}
\affiliation{Physics Department, Boston College, Chestnut Hill, Massachusetts 02467, USA}


\author{Kevin S. Bedell}
\affiliation{Physics Department, Boston College, Chestnut Hill, Massachusetts 02467, USA}


\date{\today}

\begin{abstract}
  We study the spin transport in the low temperature regime (often referred to as the precession-dominated regime) between a ferromagnetic Fermi liquid (FFL) and a normal metal metallic Fermi liquid (NFL), also known as the F/N junction, which is considered as one of the most basic spintronic devices. In particular, we explore the propagation of spin waves and transport of magnetization through the interface of the F/N junction where nonequilibrium spin polarization is created on the normal metal side of the junction by electrical spin injection. We calculate the probable spin wave modes in the precession-dominated regime on both sides of the junction especially on the NFL side where the system is out of equilibrium. Proper boundary conditions at the interface are introduced to establish the transport of the spin properties through the F/N junction. A possible transmission conduction electron spin resonance (CESR) experiment is suggested on the F/N junction to see if the predicted spin wave modes could indeed propagate through the junction. Potential applications based on this novel spin transport feature of the F/N junction are proposed in the end.
\end{abstract}
\maketitle

\section{Introduction}\label{introduction}
With the development of microelectronic devices based on electric charge reaching to its full capacity in the foreseeable future as the size of device features approaches the dimension of atoms, investigators have been eager to seek device applications based on electron spin, which has led to the emergence of a new research field called spintronics \cite{DasSarma}. The central theme of spintronics involves active manipulation of the spin degree of freedom in solid-state systems, which generally requires the generation and control of nonequilibrium spin. Over the past two decades, extensive studies on spintronics have been carried out in various solid-state systems \cite{Zutic_Fabian_DasSarma}. Among the many interesting spintronic systems, the ferromagnetic/normal metal (F/N) junction is considered to be one of the simplest and most basic, where nonequilibrium spin polarization could be generated through electrical spin injection \cite{Aronov,Johnson_Silsbee}. A considerable amount of work has been done studying the spin transport from the ferromagnetic metal to a normal metal in the classical diffusion dominated transport regime \cite{Zutic_Fabian_DasSarma}. In this paper, we investigate the spin transport through the F/N junction under electrical spin injection in the low temperature regime, where the spin diffusion is dominated by spin precession rather than collision in the classical diffusion dominated transport regime.

The relative importance of the two mechanisms has been studied both in weak ferromagnetic systems and nonequilibrium paramagnetic systems through calculating the effective spin diffusion coefficient \cite{Dahal_Bedell}. Here, we focus on studying the propagation of spin waves and transport of magnetization through the interface of the F/N junction. To be more specific, we calculate the possible transverse spin wave modes in the ferromagnetic metal side and the spin-polarized nonequilibrium normal metal side of the F/N junction using Laudau Fermi-liquid theory. We then propose a proper set of boundary conditions at the junction interface, under which the spin waves can successfully propagate from the ferromagnetic side of the F/N junction to the normal metal side. Such a phenomena could in principle be tested by a transmission conduction electron spin resonance (CESR) experiment performed on the F/N junction, and likely experimental results are discussed as well. Potential device applications based on this novel spin transport feature of the F/N junction are proposed in the end.

\section{Derivation of Spin wave modes} \label{spin_wave_modes}
\begin{figure}
  \centering
  \includegraphics[scale=0.40]{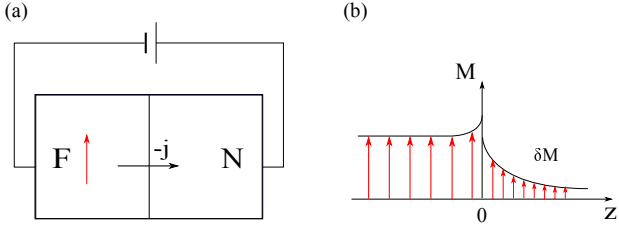}
  \caption{Pedagogical illustration of electrical spin injection into the F/N junction \cite{Zutic_Fabian_DasSarma}. (a) Schematic experimental setup; (b) Distribution of the equilibrium and nonequilibrium magnetization along $\textbf{z}$ direction (the direction of the charge current).}\label{fig:FN}
\end{figure}

Under electrical spin injection, net magnetization is driven from the ferromagnet into the normal metal region of the F/N junction by a spin-polarized charge current flowing across the F/N junction, as shown in Fig. \ref{fig:FN}(a). For a long enough relaxation time, $T_1$, of the polarized spin, this would lead to a steady state in the normal metal region of the F/N junction with nonequilibrium magnetization $\delta M$, depicted in Fig. \ref{fig:FN}(b), which we will, from here on, refer to as the spin-polarized quasiequilibrium state (QEQ) \cite{Bedell_Dahal} in the weak polarization limit, i.e., $\delta M\ll1$. Therefore, in the steady state, the F/N junction could be thought of as a composition of spin-polarized equilibrium (ferromegnet side) and quasiequilibrium (normal metal side) system. We study the transverse spin wave modes that may arise in these systems, when a small transverse spin perturbation is introduced to the steady state.

Using Landau Fermi-liquid theory, spin waves for a paramagnetic Fermi liquid in the presence of a constant external magnetic field have been well understood by solving the spin kinetic equation \cite{Baym_Pethick,Bedell_Meltzer}. These spin wave modes are the well known Silin modes \cite{Silin} for polarized Fermi liquids. A recent work has extended the study of spin waves to QEQ spin systems \cite{Bedell_Dahal}, where new gapless spin wave modes were found in a spin-polarized QEQ Fermi liquid in the absence of an external magnetic field, similar to the case of a weak ferromagnetic system.  Following the same recipe as Ref. \cite{Bedell_Dahal}, we start with the study of the spin wave modes for the QEQ state of the normal metal region of the F/N junction.

\subsection{Quasiequilibrium Fermi liquid}
Assuming the QEQ Fermi liquid having a particle density, $n=n_{\uparrow}+n_{\downarrow}$, and a small magnetization, $\sigma=n_{\uparrow}-n_{\downarrow}$, where $n_{\uparrow}, n_{\downarrow}$ are the densities of $\uparrow$, $\downarrow$ spin fermions and $\bm{\sigma}$ is polarized in an arbitrary direction, with $n=n_{\scriptscriptstyle{\text{QEQ}}}$ and $\bm{\sigma}=\bm{\sigma}^{\scriptscriptstyle{\text{QEQ}}}$ in the steady state, the kinetic equation for the spin density can be written as \cite{Baym_Pethick}
\begin{align}
\frac{\partial\bm{\sigma}_\textbf{p}}{\partial t} & +\frac{\partial}{\partial r_i}\left(\frac{\partial \varepsilon_\textbf{p}}{\partial p_i}\bm{\sigma}_\textbf{p}+\frac{\partial \textbf{h}_\textbf{p}}{\partial p_i}n_\textbf{p}\right)\nonumber \\
& +\frac{\partial}{\partial p_i}\left(-\frac{\partial \varepsilon_\textbf{p}}{\partial r_i}\bm{\sigma}_\textbf{p}-\frac{\partial \textbf{h}_\textbf{p}}{\partial r_i}n_\textbf{p}\right)\nonumber \\
& =-\frac{2}{\hbar}\bm{\sigma}_\textbf{p}\times\textbf{h}_\textbf{p}+\left(\frac{\partial\bm{\sigma}_\textbf{p}}{\partial t}\right)_\text{collision},\label{eq:spin_kinetic_equation}
\end{align}
where $\textbf{h}_\textbf{p}=\frac{2}{V}\sum_{\textbf{p}^\prime}f_{\textbf{p}\textbf{p}^\prime}^a\bm{\sigma}_{\textbf{p}^\prime}$ is the effective field taking into account only the internal field in the absence of an external magnetic field, $f_{\textbf{p}\textbf{p}^\prime}^a$ denotes the spin antisymmetric Landau Fermi-liquid interaction, $\varepsilon_\textbf{p}$ is the quasiparticle energy and $\bm{\sigma}_\textbf{p}$ is the quasiparticle spin density defined as $\bm{\sigma}_\textbf{p}\equiv-\frac{\partial n_\textbf{p}^0}{\partial \varepsilon_\textbf{p}}\frac{\bm{\sigma}}{N(0)}$, where $n_\textbf{p}^0$ is the ground state quasiparticle density distribution function (Fermi distribution function) and $N(0)=\frac{m^* p_f}{\pi^2\hbar^3}$ is the density of state at the Fermi surface. The Lorentz force term appearing in the kinetic equations of charged Fermi systems \cite{Pines_Nozieres} vanishes here as there is no external field. According to Landau Fermi-liquid theory, the spin current is given by
\begin{equation}
j_{\bm{\sigma},i}(\textbf{r},t)= 2\int \frac{d^3 p}{(2\pi\hbar)^3}\left[\frac{\partial \varepsilon_\textbf{p}}{\partial p_i}\bm{\sigma}_\textbf{p}+\frac{\partial \textbf{h}_\textbf{p}}{\partial p_i}n_\textbf{p}\right],\label{eq:spin_current}
\end{equation}
which represents the current in the $i^{\text{th}}$ spatial direction of the $\bm{\sigma}$ spin polarization. When a charge current, $\textbf{J}$, is running across the F/N junction in the steady state, we define in the QEQ Fermi liquid an average drift velocity of electrons, $\textbf{V}_0^{\scriptscriptstyle{\text{QEQ}}}$, as
\begin{equation}
\textbf{J}=-en_{\scriptscriptstyle{\text{QEQ}}}\textbf{V}_0^{\scriptscriptstyle{\text{QEQ}}}.\label{eq:3}
\end{equation}
The steady state quasiparticle velocity in the QEQ state could be approximated as
\begin{equation}
v_{\textbf{p}i}^{\scriptscriptstyle{\text{QEQ}}}= V_{0i}^{\scriptscriptstyle{\text{QEQ}}}+v_{\textbf{p}i}^0=\frac{\partial \varepsilon_{\textbf{p}}}{\partial p_i},\label{eq:particle_velocity}
\end{equation}
where $v_{\textbf{p}i}^0$ is the equilibrium quasiparticle velocity for an isotropic Fermi liquid. Substitute the QEQ quasiparticle velocity into Eq. (\ref{eq:spin_current}), we have the QEQ spin current as
\begin{align}
j_{\bm{\sigma},i}^{\scriptscriptstyle{\text{QEQ}}}(\textbf{r},t) & =
V_{0i}^{\scriptscriptstyle{\text{QEQ}}}\bm{\sigma}(1+F_0^a) \nonumber \\
& +2\int \frac{d^3 p}{(2\pi\hbar)^3} v_{\textbf{p}i}^0\bm{\sigma}_{\textbf{p}}(1+\frac{F_1^a}{3}), \label{eq:total_spin_current}
\end{align}
where we denote the first term as the drift spin current,
\begin{equation}
j_{\bm{\sigma},i}^{\scriptscriptstyle{\text{drift}}}(\textbf{r},t)= V_{0i}^{\scriptscriptstyle{\text{QEQ}}}\bm{\sigma}(1+F_0^a), \label{eq:drift_spin_current}
\end{equation}
and the second term as the regular Fermi-liquid diffusive spin current,
\begin{equation}
j_{\bm{\sigma},i}^{\scriptscriptstyle{\text{diff}}}(\textbf{r},t)=2\int \frac{d^3 p}{(2\pi\hbar)^3} v_{\textbf{p}i}^0\bm{\sigma}_{\textbf{p}}(1+\frac{F_1^a}{3}). \label{eq:diff_spin_current}
\end{equation}
In the steady state, $j_{\bm{\sigma},i}^{\scriptscriptstyle{\text{diff}}}=-D \nabla_i \bm{\sigma}^{\scriptscriptstyle{\text{QEQ}}}$, where,  $D=\frac{1}{3}v_f^2(1+F_0^a)\tau_D$, is the regular spin diffusion coefficient in Fermi-liquid theory, $v_f$ is the Fermi velocity and $\tau_D$ is the spin diffusion life time. The decomposition of the spin current into a drift term and a diffusive term is consistent with the treatment of an earlier study on the electrical spin injection into semiconductors \cite{Aronov_Pikus}.

Under a small transverse spin distortion, $\delta\bm{\sigma}$, we define the complex variables, $\sigma^\pm=\delta\sigma_x \pm i\delta\sigma_y$, and,  $j_{\bm{\sigma},i}^{\scriptscriptstyle{\text{diff}}\pm}(\textbf{r},t)=2\int \frac{d^3 p}{(2\pi\hbar)^3} v_{\textbf{p}i}^0\sigma_{\textbf{p}}^\pm(1+\frac{F_1^a}{3})$, with $\sigma_{\textbf{p}}^\pm=-\frac{\partial n_\textbf{p}^0}{\partial \varepsilon_\textbf{p}}\frac{\sigma^\pm}{N(0)}$. The oscillations of the transverse spin distortion is governed by the linearized spin conservation law derived from linearizing and summing both sides of Eq. (\ref{eq:spin_kinetic_equation}) over $\textbf{p}$,
\begin{equation}
\frac{\partial \sigma^\pm(\textbf{r},t)}{\partial t}+\frac{\partial}{\partial r_i} j_{\bm{\sigma},i}^{\scriptscriptstyle{\text{diff}}\pm}(\textbf{r},t)+ \frac{\partial}{\partial r_i} V_{0i}^{\scriptscriptstyle{\text{QEQ}}}(1+F_0^a)\sigma^\pm(\textbf{r},t)=0, \label{eq:linearized_spin_equation}
\end{equation}
where, here and through out the paper, we have assumed a very large spin relaxation time $T_1$, therefore the spin relaxation term is not included. The linearized equation of motion for the diffusive spin current takes a more complex form as
\begin{align}
\frac{\partial j_{\bm{\sigma},i}^{\scriptscriptstyle{\text{diff}}\pm}(\textbf{r},t)}{\partial t}
&+\frac{1}{3}(1+F_0^a)(1+\frac{F_1^a}{3})v_f^2\frac{\partial}{\partial r_i}\sigma^\pm(\textbf{r},t)\nonumber \\
& +(1+\frac{F_1^a}{3}) (V_{0k}^{\scriptscriptstyle{\text{QEQ}}}\frac{\partial}{\partial r_k}) j_{\bm{\sigma},i}^{\scriptscriptstyle{\text{diff}}\pm}(\textbf{r},t) \nonumber \\ &=\pm i\frac{2}{\hbar}(f_0^a-\frac{f_1^a}{3}) j_{\bm{\sigma},i}^{\scriptscriptstyle{\text{diff}}\pm}(\textbf{r},t)\sigma^{\scriptscriptstyle{\text{QEQ}}} \nonumber \\
&\quad -(1+\frac{F_1^a}{3}) j_{\bm{\sigma},i}^{\scriptscriptstyle{\text{diff}}\pm}(\textbf{r},t)/\tau_D, \label{eq:linearized_current_equation}
\end{align}
where, $F_0^a=N(0)f_0^a$ and $F_1^a=N(0)f_1^a$, are the spin antisymmetric Landau parameters, and an extra spatial gradient term on the spin current is present due to the effect of the drift charge current. Eq. (\ref{eq:linearized_spin_equation}) and Eq. (\ref{eq:linearized_current_equation}) constitute the hydrodynamic equations for the spin. After expanding, $\sigma^\pm(\textbf{r},t)$ and $j_{\bm{\sigma},i}^{\scriptscriptstyle{\text{diff}}\pm}(\textbf{r},t)$, in their corresponding Fourier series as, $\sigma^\pm(\textbf{r},t)= \int d^3q d\omega\,\sigma^\pm(\textbf{q},\omega)e^{i(\textbf{q}\cdot \textbf{r}-\omega t)}$ and $j_{\bm{\sigma},i}^{\scriptscriptstyle{\text{diff}}\pm}(\textbf{r},t)= \int d^3q d\omega \,j_{\bm{\sigma},i}^{\scriptscriptstyle{\text{diff}}\pm}(\textbf{q},\omega)e^{i(\textbf{q}\cdot \textbf{r}-\omega t)}$, respectively, the Fourier transformed hydrodynamic equations lead to a single equation for the dispersion relation,
\begin{align}
\omega^2 &+\left[\omega_1^\pm-(1+F_0^a+1+\frac{F_1^a}{3})(\textbf{V}_0^{\scriptscriptstyle{\text{QEQ}}}\cdot\textbf{q}) \right]\omega \nonumber \\ &-i(1+F_0^a)\omega_1^\pm(\textbf{V}_0^{\scriptscriptstyle{\text{QEQ}}}\cdot\textbf{q})-c_s^2 q^2 \nonumber \\ &+(1+F_0^a)(1+\frac{F_1^a}{3})|\textbf{V}_0^{\scriptscriptstyle{\text{QEQ}}}\cdot\textbf{q}|^2=0,
\label{eq:equation_dispersion}
\end{align}
where $\omega_1^\pm=i\left[(1+\frac{F_1^a}{3})/\tau_D\mp i\frac{2}{\hbar} (f_0^a-\frac{f_1^a}{3})\sigma^{\scriptscriptstyle{\text{QEQ}}}\right]$, and $c_s^2=\frac{1}{3}(1+F_0^a)(1+\frac{F_1^a}{3})v_f^2$. We solve Eq. (\ref{eq:equation_dispersion}) in the long wave length, small $\textbf{q}$, limit, where we keep only terms of order $q^2$ and below. The dispersion relations of the modes are given as
\begin{subequations}
\label{eq:QEQ_dispersions}
\begin{align}
&\omega_0^\pm(q)= (1+F_0^a)(\textbf{V}_0^{\scriptscriptstyle{\text{QEQ}}}\cdot\textbf{q}) - iD_{\text{eff}}^\pm q^2, \label{eq:QEQ_dispersion_0}\\
&\omega_1^\pm(q)=-\omega_1^\pm+(1+\frac{F_1^a}{3})(\textbf{V}_0^{\scriptscriptstyle{\text{QEQ}}} \cdot\textbf{q})+ iD_{\text{eff}}^\pm q^2, \label{eq:QEQ_dispersion_1}
\end{align}
\end{subequations}
where, $D_{\text{eff}}^\pm=ic_s^2/\omega_1^\pm$, is interpreted as the effective spin diffusion coefficient. The meaning of $D_{\text{eff}}^\pm$ becomes clearer when Eq. (\ref{eq:linearized_current_equation}) is rearranged under the steady state condition, $\partial j_{\bm{\sigma},i}^{\scriptscriptstyle{\text{diff}}\pm}(\textbf{r},t)/\partial t=0$,
\begin{align}
j_{\bm{\sigma},i}^{\scriptscriptstyle{\text{diff}}\pm}(\textbf{r},t)= &-D_{\text{eff}}^\pm \frac{\partial}{\partial r_i}\sigma^\pm(\textbf{r},t) \nonumber \\
&-\frac{D_{\text{eff}}^\pm}{v_f^2(1+F_0^a)/3} (V_{0k}^{\scriptscriptstyle{\text{QEQ}}}\frac{\partial}{\partial r_k}) j_{\bm{\sigma},i}^{\scriptscriptstyle{\text{diff}}\pm}(\textbf{r},t). \label{eq:spin_diffusion}
\end{align}
For small $\textbf{q}$, the second term on the right hand side of Eq. (\ref{eq:spin_diffusion}) is an order of $q$ higher than the first term, we can thus drop the last term in Eq. (\ref{eq:spin_diffusion}) and recover the familiar Fick form \cite{Baym_Pethick} for the spin current,
\begin{equation}
j_{\bm{\sigma},i}^{\scriptscriptstyle{\text{diff}}\pm}(\textbf{r},t)= -D_{\text{eff}}^\pm \frac{\partial}{\partial r_i}\sigma^\pm(\textbf{r},t), \label{eq:Fick_form}
\end{equation}
where $D_{\text{eff}}^\pm$ clearly serves as the role of the effective spin diffusion coefficient.

The spin precession term, $\pm\frac{2}{\hbar} (f_0^a-\frac{f_1^a}{3})\sigma^{\scriptscriptstyle{\text{QEQ}}}$, in the denominator of $D_{\text{eff}}^\pm$, often referred to as the Leggett-Rice effect \cite{Leggett_Rice,Legget}, is solely a consequence of the interactions between quasiparticles; it would cease to exist had we treated the electronic system in the normal metal as a free Fermi liquid using the simple electron band structure model. The effective spin diffusion coefficient reduces to the regular Fermi liquid spin diffusion coefficient, $D=\frac{1}{3}v_f^2(1+F_0^a)\tau_D$, if we shut off $F_l^a$'s for the QEQ system. The complete picture of the competition between the collision effect and the spin precession effect in the effective spin diffusion coefficient over a wide temperature range was obtained experimentally in liquid $^3$He \cite{Corruccini_Osheroff_Lee_Richardson}, and was studied theoretically in the spin polarized Fermi liquids \cite{Dahal_Bedell} as well.

In the low temperature limit, we take the spin diffusion lifetime $\tau_D\approx \infty$, since it varies as $T^{-2}$ in a clean Fermi liquid, therefore, the collision term becomes negligible in $D_{\text{eff}}^\pm$ compared to the spin precession term, leading to a purely imaginary effective spin diffusion coefficient. Finally, the dispersion relations of the modes for the QEQ system in the low temperature precession dominated regime can be expressed as
\begin{subequations}
\label{eq:QEQ_dispersions_low_T}
\begin{align}
\omega_0^\pm(q)= & \,(1+F_0^a)(\textbf{V}_0^{\scriptscriptstyle{\text{QEQ}}}\cdot\textbf{q}) \pm \frac{c_s^2 q^2}{\frac{2}{\hbar}(f_0^a-\frac{f_1^a}{3})\sigma^{\scriptscriptstyle{\text{QEQ}}}}, \label{eq:QEQ_dispersion_0_low_T}\\
\omega_1^\pm(q)= & (1+\frac{F_1^a}{3})(\textbf{V}_0^{\scriptscriptstyle{\text{QEQ}}} \cdot\textbf{q}) \mp \frac{c_s^2 q^2}{\frac{2}{\hbar}(f_0^a-\frac{f_1^a}{3})\sigma^{\scriptscriptstyle{\text{QEQ}}}} \nonumber \\
&\mp\frac{2}{\hbar}(f_0^a-\frac{f_1^a}{3})\sigma^{\scriptscriptstyle{\text{QEQ}}}. \label{eq:QEQ_dispersion_1_low_T}
\end{align}
\end{subequations}
Since both dispersion relations of the modes contain only real terms, for small enough $\textbf{q}$, we have found the transverse spin wave modes that survive from Landau damping \cite{Baym_Pethick} and can propagate through the QEQ system, which serve as collective excitations of the QEQ system.

\subsection{Ferromagnetic Fermi liquid}
We consider here a weak ferromagnetic Fermi liquid for the ferromagnet region of the F/N junction, where the spin dynamics could be studied in the language of Landau Fermi-liquid theory \cite{Bedell_Blagoev} in a similar fashion as the QEQ system. We should use $\bar{F}_l^a$ and $\bar{f}_l^a$ for the Fermi liquid parameters in the ferromagnetic metal side of the F/N junction. In the precession dominated regime, the transverse spin wave modes in the ferromagnet region of the F/N junction turn out to be nearly identical to the ones found in the normal metal region of the F/N junction, i.e., the QEQ Fermi liquid,
\begin{subequations}
\label{eq:FM_dispersions}
\begin{align}
\omega_0^\pm(q)= & \,(1+\bar{F}_0^a)(\textbf{V}_0^{\scriptscriptstyle{\text{FM}}}\cdot\textbf{q}) \mp \frac{\bar{c}_s^2 q^2}{\frac{2}{\hbar}(\bar{f}_0^a-\frac{\bar{f}_1^a}{3})\sigma^{\scriptscriptstyle{\text{FM}}}}, \label{eq:FM_dispersion_0}\\
\omega_1^\pm(q)= & (1+\frac{\bar{F}_1^a}{3})(\textbf{V}_0^{\scriptscriptstyle{\text{FM}}} \cdot\textbf{q}) \pm \frac{\bar{c}_s^2 q^2}{\frac{2}{\hbar}(\bar{f}_0^a-\frac{\bar{f}_1^a}{3})\sigma^{\scriptscriptstyle{\text{FM}}}} \nonumber \\
&\mp\frac{2}{\hbar}(\bar{f}_0^a-\frac{\bar{f}_1^a}{3})\sigma^{\scriptscriptstyle{\text{FM}}}, \label{eq:FM_dispersion_1}
\end{align}
\end{subequations}
except for the definition of the spin wave velocity, $\bar{c}_s^2=\frac{1}{3}|1+\bar{F}_0^a|(1+\frac{\bar{F}_1^a}{3})\bar{v}_f^2$, since, $(1+\bar{F}_0^a)<0$, for a ferromagnetic Fermi liquid, $\bar{v}_f$ is the Fermi velocity of the electrons in the ferromagnet, $\sigma^{\scriptscriptstyle{\text{FM}}}$ is the equilibrium spin polarization in the ferromagnet, and $\textbf{V}_0^{\scriptscriptstyle{\text{FM}}}$ is the drift velocity of electrons in the ferromagnet, which is related to the charge current through, $\textbf{J}=-en_{\scriptscriptstyle{\text{FM}}} \textbf{V}_0^{\scriptscriptstyle{\text{FM}}}$, with $n_{\scriptscriptstyle{\text{FM}}}$ being the equilibrium carrier density in the ferromagnet. Again, for small $\textbf{q}$, the transverse spin wave modes represented by Eq. (\ref{eq:FM_dispersions}) are the propagating modes in the ferromagnet region of the F/N junction.

\subsection{Boundary conditions} \label{boundary_condition}
So far, we have established the propagating transverse spin wave modes in the ferromagnet region and the normal metal region of the F/N junction, respectively. Naturally, one would want to look for proper boundary conditions to make the spin wave modes propagate through the interface of the F/N junction, as it could greatly increase the functionality of the F/N junction as a spintronic device. To simplify the analysis while keeping the underlying physics unchanged, we treat the F/N junction as an effectively one dimensional structure, where only the spatial variation in the $\textbf{z}$ direction of the spin density is non-zero.

In describing the steady state of the F/N junction under electrical spin injection, we have adopted the boundary conditions analogous to earlier studies on spin injection into metals \cite{Aronov} and semiconductors \cite{Aronov_Pikus}, where the total spin current is continuous at the interface in the absence of surface spin relaxation,
\begin{equation}
j_{\bm{\sigma},z}^{\scriptscriptstyle{\text{FM}}}(z,t)= j_{\bm{\sigma},z}^{\scriptscriptstyle{\text{QEQ}}}(z,t) \quad \text{for} \;z=0. \label{eq:continuous_current}
\end{equation}
Contrary to the steady state spin polarization, we assume hard boundary conditions on the oscillations of the small transverse spin distortion,
\begin{subequations}
\label{eq:distortion_boundary}
\begin{align}
\frac{\partial \sigma_{\scriptscriptstyle{\text{FM}}}^\pm}{\partial z}&=0\quad\text{for}\;\; z=-L_1\;\text{and}\;0, \label{eq:distortion_boundary_FM} \\
\frac{\partial \sigma_{\scriptscriptstyle{\text{QEQ}}}^\pm}{\partial z}&=0\quad\text{for}\;\; z=0\;\text{and}\;L_2. \label{eq:distortion_boundary_QEQ}
\end{align}
\end{subequations}
In other words, the diffusive spin current from distortion vanishes at the surfaces, $j_{\bm{\sigma},z}^{\scriptscriptstyle{\text{diff}}\pm}(z,t)= -D_{\text{eff}}^\pm \frac{\partial \sigma^\pm}{\partial z}=0$ for $z=0,-L_1\;\text{and}\;L_2$, where, $L_1$ and $L_2$, are the widths of the ferromagnet region and the normal metal region of the F/N junction, respectively. Under the conditions of Eq. (\ref{eq:distortion_boundary_FM}), a series of standing wave modes with wave numbers, $q_z=n\pi/L_1$, could be excited for the transverse spin waves in the ferromagnet region of the F/N junction. These standing wave modes should appear as sidebands on the electron spin-resonance line  analogous to the spin wave excitations in nonferromagnetic metals in transmission CESR experiments \cite{Platzman_Wolff,Schultz_Dunifer}, which we will discuss to some extent in a later section of this paper. Under spin wave excitation, transverse spin oscillations are transmitted from the left edge of the ferromagnet region to the interface of the F/N junction through the propagation of the standing wave modes, resulting in the accumulation of oscillating transverse spin signals at the interface. Recalling that we have required the total spin current to be continuous, as well as the vanishing of the diffusive spin current, $j_{\bm{\sigma},z}^{\scriptscriptstyle{\text{diff}}\pm}(z,t)=0$, at the interface, the drift spin current must then be continuous at the interface, $j_{\scriptscriptstyle{\text{FM}},\bm{\sigma},z}^{\scriptscriptstyle{\text{drift}}\pm}(z,t) =j_{\scriptscriptstyle{\text{QEQ}},\bm{\sigma},z}^{\scriptscriptstyle{\text{drift}}\pm}(z,t)$ for $z=0$, which leads to the following relation,
\begin{equation}
V_{0z}^{\scriptscriptstyle{\text{FM}}}\sigma_{\scriptscriptstyle{\text{FM}}}^\pm(1+\bar{F}_0^a) =V_{0z}^{\scriptscriptstyle{\text{QEQ}}}\sigma_{\scriptscriptstyle{\text{QEQ}}}^\pm(1+F_0^a) \quad \text{for}\;\;z=0. \label{eq:drift_continuous}
\end{equation}
The transverse spin signals in the ferromagnet side of the interface are driven into the normal metal region by the external electric potential in the form of a continuous drift spin current. Consequently, propagating transverse spin wave modes are excited in the normal metal region of the F/N junction once the transverse spin distortion is driven into the normal metal region. Therefore, Eq. (\ref{eq:continuous_current}) and Eq. (\ref{eq:distortion_boundary}) constitute the boundary conditions under which the spin wave modes can effectively propagate across the F/N junction.

\section{Result and discussion} \label{results_discussion}
\subsection{Spin wave modes}
\begin{figure}[t]
  \centering
  \includegraphics[scale=0.38]{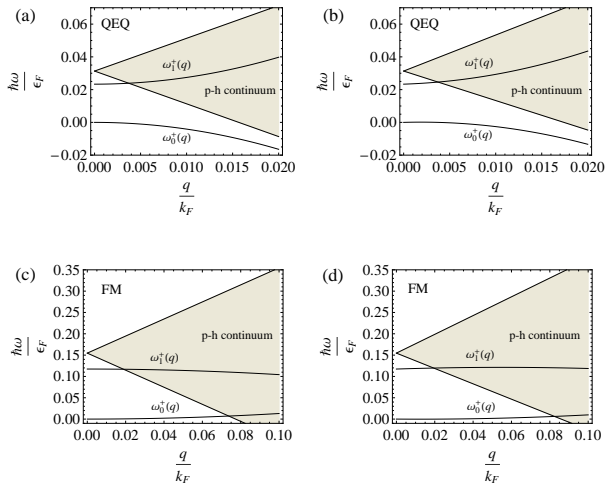}
  \caption{Dispersion relations of the spin wave modes and p-h continuum of the QEQ system and Ferromagnetic system. (a) QEQ system with $F_0^a=-0.235$, $F_1^a=-0.18$, $V_0^{\scriptscriptstyle{\text{QEQ}}}=0$; (b) QEQ system with $F_0^a=-0.235$, $F_1^a=-0.18$, $V_0^{\scriptscriptstyle{\text{QEQ}}}/v_f=10\%$; (c) Ferromagnetic system with $\bar{F}_0^a=-1.16$, $\bar{F}_1^a=-0.84$, $V_0^{\scriptscriptstyle{\text{FM}}}=0$; (d) Ferromagnetic system with $\bar{F}_0^a=-1.16$, $\bar{F}_1^a=-0.84$, $V_0^{\scriptscriptstyle{\text{FM}}}/\bar{v}_f=10\%$.} \label{fig:waves}
\end{figure}
Aside from the collective modes developed in the previous sections for both QEQ and ferromagnetic systems, there are also a continuum of particle-hole (p-h) excitations in these systems. The dispersions of the p-h excitations for the QEQ system are \cite{Bedell_Dahal}, $\omega_{\text{p-h}}^{\pm}(q)=\mp \frac{2}{\hbar}\sigma^{\scriptscriptstyle{\text{QEQ}}} f_0^a + \textbf{q}\cdot \textbf{v}_{\textbf{p}}$, where, $\textbf{v}_{\textbf{p}}=\textbf{V}_0^{\scriptscriptstyle{\text{QEQ}}}+\textbf{v}_{\textbf{p}}^0$, is the steady state quasiparticle velocity given by Eq. (\ref{eq:particle_velocity}), $|\textbf{v}_{\textbf{p}}^0|=v_f$, and the dispersions of the p-h continuum for the ferromagnetic systems are \cite{Zhang_Farinas_Bedell}, $\omega_{\text{p-h}}^{\pm}(q)=\mp \frac{2}{\hbar}\sigma^{\scriptscriptstyle{\text{FM}}} \bar{f}_0^a + \textbf{q}\cdot \bar{\textbf{v}}_{\textbf{p}}$, where, $\bar{\textbf{v}}_{\textbf{p}}=\textbf{V}_0^{\scriptscriptstyle{\text{FM}}}+\bar{\textbf{v}}_{\textbf{p}}^0$, and, $|\bar{\textbf{v}}_{\textbf{p}}^0|=\bar{v}_f$. For a given $\textbf{q}$, with the freedom of choosing $\textbf{v}_{\textbf{p}}$ over the entire Fermi surface, the dispersions of the p-h excitations form a continuum bounded by the maximum and minimum values of $\textbf{q}\cdot\textbf{v}_{\textbf{p}}$. Here, we choose $\textbf{q}$, $\textbf{V}_0^{\scriptscriptstyle{\text{QEQ}}}$ and $\textbf{V}_0^{\scriptscriptstyle{\text{FM}}}$ to be in the $\textbf{z}$ direction in displacing the dispersion relations of the spin wave modes, as we consider the F/N junction an effectively one dimensional system. The dispersions of the spin wave modes for both the QEQ and ferromagnetic systems together with their respective p-h continuums are plotted in Fig. \ref{fig:waves}, where we have chosen a spin polarization of $10\%$, $\sigma/n=10\%$, for both systems. We have chosen to show only the upper branches $\omega^+(q)$ of the spin wave modes, as the physics is the same for the two branches except for the direction of the spin precession. We use the set of Landau parameters suitable for the simple metal aluminum \cite{Rice} when evaluating the dispersion relations for the QEQ system on the normal metal region of the F/N junction. The Landau parameters for the ferromagnetic system are obtained from the weakly ferromagnetic heavy fermion material MnSi \cite{Zhang_Farinas_Bedell}. In a typical electrical spin injection experiment on the F/N junction \cite{Johnson_Silsbee} composed of simple metals nickel and iron on the ferromagnet side and aluminum on the normal metal side, the electron drift velocity $\textbf{V}_0$ is negligible compared with the Fermi velocity for a small driven current of order $10\sim 100\,\text{mA}$, however $\textbf{V}_0$ can be appreciable with respect to the Fermi velocity when we use a heavy fermion material with a big effective mass, which as a consequence reduces the Fermi velocity. In fact, for an F/N junction with a cross section of $1\mu m\times 1 \mu m$ and a driven current of $1\,\text{A}$, we could have the drift velocity as close as $10\%$ of the Fermi velocity, $\text{V}_0/v_f\approx 10\%$, if the effective mass $m^*$ of the heavy fermion material is of order $10^2\sim 10^3 \, m_e$ and the electron density is that of a typical metal, $n\approx 10^{23}cm^{-3}$. Hence, the spin wave dispersions for both the zero drift velocity case and the $10\%$ drift velocity case are shown in Fig. \ref{fig:waves}.

As is shown in Fig. \ref{fig:waves}(a) and \ref{fig:waves}(b), there exists a gapless mode as well as a gapped spin wave mode in the QEQ system despite the absence of an external magnetic field. The accumulation of the nonequilibrium spin polarization in the QEQ system has effectively broken the spin symmetry in the original paramagnetic state, therefore making it possible for the existence of these collective excitations. Although the dispersion relation of the gapless spin wave mode in the QEQ system is very similar to the gapless Goldstone mode of the ferromagnetic system shown in Fig. \ref{fig:waves}(c) and \ref{fig:waves}(d), their respective origins are fundamentally different as has been discussed in detail in Ref. \cite{Bedell_Dahal}. The gapless spin wave modes in both the QEQ and the ferromagnetic systems are related to the uniform precessional mode of the spin polarization. This is understood through the following argument. If we take $\textbf{q}=0$, each individual spin is polarized in the same direction, hence no uniform precession of the individual spins around the internal field will take place, and we have $\omega(0)=0$, which resembles a gapless energy spectrum for the uniform spin precessional mode. Consequently, the gapped spin wave modes must be related to the precessional mode of the spin current, as the two spin wave modes are the solutions to the coupled hydrodynamic equations of the spin polarization, Eq. (\ref{eq:linearized_spin_equation}), and the spin current, Eq. (\ref{eq:linearized_current_equation}). The gapped modes are collective excitations of the system which involve energy consuming spin flip processes, and could also be interpreted as the Higgs amplitude mode in a weak ferromagnet \cite{Zhang_Farinas_Bedell}. For a small enough $\textbf{q}$, the dispersion curves of the spin wave modes are outside the p-h continuums, therefore the collective excitations become propagating spin wave modes without getting Landau damped. It has to be pointed out that although the gapless modes seem to survive entirely from the p-h continuum, the calculation is only accurate in the low $\textbf{q}$ limit. Corrections to the dispersion relations as well as the p-h continuums at higher $\textbf{q}$ need to be evaluated through calculating the complete spin response function of the system, which is beyond the scope of this paper.

\subsection{Transmission CESR experiment}
The transmission CESR experiment has long been used in investigating the spin wave excitations in paramagnetic metals \cite{Platzman_Wolff,Schultz_Dunifer}, where, by coupling a microwave power to one side of the metal sample, the spin wave modes are excited in the sample and spin signals are transmitted through the sample to be detected by the receiver on the other side. When the frequency of the incident microwave power satisfies the condition, $\omega=\omega(q)$ with $q=n\pi/L$, where $\omega(q)$ is the dispersion of the spin wave modes and $L$ is the width of the sample, the system is under spin resonance with standing spin wave modes being excited in the sample and there appears a peak in the intensity of the transmitted spin signals collected by the receiver. A typical set of transmitted spin signal data from the transmission CESR experiment would contain multiple peaks over a range of frequency.

We propose a transmission CESR experiment on the F/N junction under the condition of electrical spin injection to probe the spin wave modes calculated in section \ref{spin_wave_modes} and to test the proposal of propagating spin wave modes across the F/N junction. Instead of applying an external magnetic field to the sample and sweeping through a range of the magnetic fields during the measurement of a traditional transmission CESR experiment, we propose not to apply any external magnetic field to the sample, but vary the frequency of the incident microwave power instead. Spin wave modes can be excited in the QEQ system without the introduction of the external magnetic field, which is a major difference between the QEQ state and the equilibrium paramagnetic state. Under the boundary conditions introduced in section \ref{boundary_condition}, electron spin resonance is achieved in the F/N junction when standing spin wave modes are excited on both sides of the F/N junction. In the absence of an external magnetic field, under electron spin resonance, the frequency of the exciting microwave power must satisfy the following condition,
\begin{equation}
\omega=\omega^{\scriptscriptstyle{\text{FM}}}(q_1)=\omega^{\scriptscriptstyle{\text{QEQ}}}(q_2),
\label{eq:resonance_condition}
\end{equation}
where, $q_1=n_1\pi/L_1$ and $q_2=n_2\pi/L_2$, are the respective wave vectors of the standing wave modes in the ferromagnetic system and the QEQ system, $\omega^{\scriptscriptstyle{\text{FM}}}(q)$ and $\omega^{\scriptscriptstyle{\text{QEQ}}}(q)$ stand for the spin wave dispersions of the ferromagnetic system and the QEQ system presented in Eq. (\ref{eq:FM_dispersions}) and Eq. (\ref{eq:QEQ_dispersions_low_T}), respectively. The spin signals measured from the transmission CESR experiment are expected to display a series of intensity peaks located at the frequencies derived from Eq. (\ref{eq:resonance_condition}) known as the spin resonance lines. Each spin resonance line represents the excitation of the standing spin wave modes on both sides of the F/N junction with a distinct pair of wave vectors $(n_1\pi/L_1, n_2\pi/L_2)$. The positions of the spin resonance lines depend on the experimental parameters such as the values of the Landau parameters of the metals forming the F/N junction, the widths of the two regions of the F/N junction, the degree of spin polarizations in the two regions of the F/N junction and the magnitude of the driving charge current.

For the particular MnSi/Al junction shown in Fig. \ref{fig:waves}, the frequencies of the spin wave modes in the small $\textbf{q}$ limit can be estimated. We find on the ferromagnet (MnSi) side of the F/N junction, $\nu_0\in(0\,\text{Hz},10^{12}\,\text{Hz})$ for the frequency of the gapless mode and $\nu_1\sim10^{13}\,\text{Hz}$ for the frequency of the gapped mode, while, $\nu_0\in(0\,\text{Hz},10^{14}\,\text{Hz})$ and $\nu_1\sim10^{14}\,\text{Hz}$, on the normal metal (Al) side. Therefore, under electron spin resonance, both the gapless and the gapped modes could be excited in MnSi, whereas only the gapless mode is excited in Al. The frequency of the gapped mode in Al is much higher than that of both spin wave modes in MnSi, so the simultaneous excitation of the gapped mode in Al and either of the spin wave modes in MnSi by a single microwave power is impossible in principle. A proper width of the sample could be chosen, $L\sim1\,\mu m$, so that the wave vectors, $q=n\pi/L\sim10^6\,m^{-1}$, of the standing spin wave modes are much smaller than the Fermi wave vector, $k_F\sim10^{10}\,m^{-1}$, and it guarantees that we are working in the long wave length limit, $q\ll k_F$, where the spin waves survive from Landau damping and become propagating modes.

\subsection{Potential applications}
According to our study, propagating spin wave modes could be excited in the normal metal side of the F/N junction without applying an external magnetic field on it. Spin wave modes excited in the ferromagnetic side of the F/N junction are also proposed to be able to propagate across the interface of the junction and travel through the normal metal side of the F/N junction. This unique spin transport feature of the F/N junction makes it possible to speculate about potential device applications. Since radio frequency signals with certain frequencies could effectively tunnel through the F/N junction, we could think of the F/N junction as a frequency selective signal transmitter. More importantly, we can dynamically control the resonance conditions of the F/N junction by varying the relevant parameters of the F/N junction. As a result, we can effectively turn the transmitter on and off with respect to a microwave signal with a particular frequency by moving the F/N junction to and away from the resonance. The F/N junction then serves as a novel switch like device in terms of its ability in transmitting microwave signals. The easiest and most practical way to control the switch is varying the drift current through changing the electric bias potential applied to the F/N junction. However, as mentioned previously, the effect of the electron drift velocity is rather negligible in an F/N junction composed of simple metals, we need the F/N junction made of heavy fermion materials in order to utilize this control mechanism. Other control mechanisms such as dynamically controlling the spin polarization in the F/N junction are also worth exploring. It is also possible to realize multiple resonance conditions through changing a single or multiple parameters of the F/N junction, therefore the F/N junction could be turned into a more functional transmitter with one or more controlling dials, which could be tuned to make the device transmit microwave signals with different frequencies.

\section{Conclusion} \label{conclusion}
We have studied the spin transport and spin dynamics in the F/N junction under electrical spin injection in the low temperature (precession dominated) regime using Landau Fermi liquid theory. In particular, we calculate the transverse spin wave modes on both sides of the F/N junction. The normal metal region of the F/N junction is treated as a QEQ system with nonequilibrium spin polarization. We find both a gapless and a gapped spin wave mode in the QEQ system similar to a weak ferromagnetic system, which makes the QEQ system fundamentally different from an equilibrium paramagnetic system. Probable propagation of the spin wave modes through the F/N interface is proposed and a transmission CESR experiment on the F/N junction is suggested to test such a proposal. If the proposal is valid, we will see multiple spin resonance lines in the transmitted spin signals from the transmission CESR experiment similar to the result of a transmission CESR experiment on a paramagnetic metal. In the end, potential device applications are speculated for the F/N junction, and we suggest that a novel switch like device as well as a functional microwave signal transmitter could be made out of the F/N junction with a couple of control mechanisms being mentioned as well.

\section{Acknowledgements}
We thank Hari Dahal, Chris Hammel and Darryl Smith for valuable and insightful discussions. This work is supported in part by John H, Rourke, Boston College endowment fund.

\end{document}